# Chapter 1
# Adaptive Traffic Fingerprinting for Darknet Threat Intelligence

Hamish Haughey, Gregory Epiphaniou, Haider Al-Khateeb, and Ali Dehghantanha

**Abstract** Darknet technology such as Tor has been used by various threat actors for organising illegal activities and data exfiltration. As such there is a case for organisations to block such traffic, or to try and identify when it is used and for what purposes. However, anonymity in cyberspace has always been a domain of conflicting interests. While it gives enough power to nefarious actors to masquerade their illegal activities, it is also the corner stone to facilitate freedom of speech and privacy. We present a proof of concept for a novel algorithm that could form the fundamental pillar of a darknet-capable Cyber Threat Intelligence platform. The solution can reduce anonymity of users of Tor, and considers the existing visibility of network traffic before optionally initiating targeted or widespread BGP interception. In combination with server HTTP response manipulation, the algorithm attempts to reduce the candidate data set to eliminate client-side traffic that is most unlikely to be responsible for server-side connections of interest. Our test results show that MITM manipulated server responses lead to expected changes received by the Tor client. Using simulation data generated by shadow, we show that the detection scheme is effective with false positive rate of 0.001, while sensitivity detecting non-targets was 0.016±0.127. Our algorithm could assist collaborating organisations willing to share their threat intelligence or cooperate during investigations.

Hamish Haughey
University of Northumbria, Newcastle, England, e-mail: `hamish.haughey@gmail.com`

Gregory Epiphaniou
University of Bedfordshire, Bedfordshire, England e-mail: `gregory.epiphaniou@beds.ac.uk`

Haider Al-Khateeb
University of Bedfordshire, Bedfordshire, England e-mail: `haider.alkhateeb@beds.ac.uk`

Ali Dehghantanha
University of Salford, Salford, England e-mail: `A.Dehghantanha@salford.ac.uk`

## 1.1 Introduction

Threats to individuals and organisations from Cyber attackers have been observed since the early days of computers and the Internet [48]. Threat actors have perpetrated various attacks ranging from relatively innocuous hoaxes to more impact-ful instances of social engineering and reverse engineering to harvest credentials, hold organisations to ransom for their data, or cause actual physical damage to systems [30]. In response to this persistent threat from a wide range of actors with varying motives and methods, a number of ontologies for Cyber Threat Intelligence (CTI) have appeared over the years such as STIX and TAXII [10], OpenIOC [8], SCAP [39], VERIS [46], Cybox [9], and RID and IODEF [25]. Given that so many ontologies exist to address some aspect of exchanging CTI, further work has attempted to taxonomise these systems to understand their dependencies and interoperability [6]. Furthermore, it is widely accepted that a critical part of an organisation's CTI capability requires the sharing of information with trusted peers [2]. This varied selection of offerings highlights the importance of standardisation, as organisations are likely to use a particular solution that may or not be compatible with that in use by another organisation. A gap in the current offerings appears to be in satisfactorily addressing Privacy Enhancing Technology (PET) as a medium for threat actors to perpetrate their nefarious activities undetected.

An example of threat actor behaviour is the use of encrypted channels for data exfiltration [20], and the challenge for organisations in identifying or blocking known bad network locations has increased due to the readiness with which PET is now available. Organisations are then faced with a choice, to allow or disallow traffic originating from, or destined for, such PET on their networks. This choice is further complicated by the open debate on privacy and data protection [13].

It is easily observable that long-standing concerns regarding privacy and anonymity continue to grow among certain groups [47, 35, 17]. This may not come as a surprise, considering the increasing capabilities of some organisations to monitor and report on user behaviour in day to day activities [14].

Fuelling this debate, there have been recent legislations such as the Investigatory Powers Bill in the UK proposed to require Internet Service Providers (ISPs) and Mobile Operators to preserve meta-data on the activities of each Internet user [56]. This has resulted in greater use over time of PET such as The Onion Router (Tor) [55, 15], The Invisible Internet Project (i2p) [24], and Freenet [52], for a wide range of reasons, and no longer limited to cyber-espionage or illegal activities.

Deep Packet Inspection (DPI) is a suitable example of a technique to extend analysis capabilities towards encrypted traffic [61]. For instance, while DPI cannot by default access encrypted streams, it could still facilitate censorship by means of a Denial-of-Service (DoS) attack on PET by analysing IP and TCP headers [61]. Countermeasures in this case focus on obfuscating PET traffic [63, 60]. For example, Tor's obfsproxy [43] is implemented to mock the behaviour of the widely used Transport Layer Security (TLS) protocol, relying on the essential role that TLS plays in other communications, and the fact that it must be permitted as a key requirement to enable e-commerce in a given region.

There is a known threat to users of PET of powerful adversaries with access to Autonomous Systems (AS) or Internet Exchanges (IX), who are in strong tactical position to view or gain access to large portions of Internet traffic, potentially allowing passive traffic association to take place [57, 51]. While the Tor threat model states that it does not protect against adversaries that can view both sides of a circuit, the Tor path selection algorithm does take steps to reduce the chance of this happening [15, 34] and therefore such an adversary is obviously of concern to the developers and users of the system.

In this work we present a novel algorithm that could act as a fundamental pillar of a Threat Intelligence Platform for use by AS and IX operators either alone, or in collaboration with trusted peers. The algorithm may allow operators to identify encrypted connections engaging in activity that is against their acceptable use policies or terms and conditions of use. The algorithm fingerprints TCP connection meta-data and supplies a fully automated routine to assist with the effective degradation of un-traceability of PET users. The proposed algorithm combines several previously documented techniques. The algorithm can classify network streams according to flow metrics, make use of BGP interception if necessary to increase the attack surface for traffic association, and also manipulate server-side traffic destined for the client. Our initial test results show that MITM manipulated server responses lead to expected changes received by the Tor client. Using simulation data generated by shadow, we show that the detection scheme is effective with false positive rate of 0.001, while sensitivity detecting non-targets was 0.016±0.127. We believe that the algorithm can be further improved or adapted in order to improve detection rates and efficiencies in performance. Our algorithm could assist collaborating organisations willing to share their threat intelligence or cooperate during investigations.

The traffic association methods alone may prove useful to routing providers that wish to engage in intelligence-sharing, by allowing a risk score to be assigned to flows that demonstrate particular behaviour. TCP flow metrics could be combined with other scoring metrics that service providers wish to use, or may highlight candidates for the traffic manipulation or BGP interception components. Such collaboration between providers could lead to improved overall risk-reduction or may prove valuable during forensics investigations.

The remainder of this chapter is structured as follows: Section 1.2 contains background and discussion of existing works describing attack and defence mechanisms relevant to our work. Then, Section 1.3 introduces an adaptive traffic association and BGP interception algorithm (ATABI) against Tor. In section 1.4 we present our initial experimentation and results and discuss them in 1.5. Finally, the chapter is concluded in Section 1.6.

## 1.2 Background

It is important for any organisation to have some handle on the common, generic, and opportunistic attacks that they may be subject to. For larger or more risk-averse organisations, there is a case for being aware of more targeted attacks, and applicable Advanced Persistent Threats (APT) [42]. Threat intelligence can be generated from a variety of resources that a typical organisation has access to, such as web server logs, firewall logs, mail logs, antivirus, and host or network intrusion detection systems [38]. The consumption and interpretation of such data can become a challenge due to the large volumes often generated, leading to a case of being unable to see the wood for the trees, or searching for a needles in a haystacks [45]. Often critical to identifying threats is the challenging task of defining the baseline and understanding what normal activity actually looks like [59]. Furthermore, threats are ever evolving, and older threats often tend to become benign as more effective techniques take over, or when law enforcement take down command and control infrastructure [41].

The primary goal of CTI, then, is to inform organisations of what the current threats are so that appropriate actions can be taken, with as much automation as possible [7]. To this end machine learning, big data analytics, and intelligence sharing techniques have become more common in modern Security and Information and Event Management (SIEM) and CTI systems [50]. Primary customers of such technology are always going to be determined by factors such as their risk profile, risk appetite, and of course the size of their security budget [7].

### 1.2.1 Analysis of Attack Vectors in Tor

Having a variety of threat intelligence sources increases an organisation's ability to identify threats early on in a typical series of sequential actions leading to data breach. Where PET such as Tor is concerned, the ability to inject relays allows an observer to become a part of the network, and when they are participating as an entrance or exit node, the source or the destination of traffic could be recorded by a threat management platform. Indeed, many attacks against Tor users require visibility of both the entrance and exit traffic [34, 57]. Injection of relays is typically simple to achieve as PET systems are mostly designed to allow anyone to participate as a user or a router of traffic [53, 64, 31, 12, 32, 54, 1, 3, 5]. As such, adversaries can easily initiate rogue routers in order to conduct active traffic analysis by injecting traffic into legitimate communication channels in an attempt to enumerate and analyse underlying traffic [31, 32].

These common and easy-to-deploy attacks may effectively expose anonymous communications, and the current authors assert that they may also give service providers and collaborating organisations the opportunity to record and share observed behaviour originating from specific locations. The more routers that are compromised by the adversary, the greater the probability that circuits will start being built using those routers as entrance and exit nodes. This comes with a financial cost increasing over time with the total number of Tor relays in existence. It is possible with this prerequisite for a Tor exit relay to insert a signal into the traffic that can be detected by an entry relay [31]. This is accomplished by making changes to the Tor application code to control at what point write events are made, which result in output buffers being flushed. By controlling whether either one or three cells [1] are sent in an individual Internet Protocol (IP) packet, it is possible to create a binary signal. If the signal created by the exit relay matches the signal received by the entry relay, then the user will be discovered along with the visited website. The obstacle that must be overcome by an adversary to perpetrate this class of attack is in getting the target to use entry and exit routers belonging to an adversary, or in gaining control of routers belonging to other operators.

Some of the most powerful attacks against PET are those possible if an adversary has access to AS or IX routing infrastructure [33, 21, 62, 34, 18, 51, 57], as this position greatly increases the possibility of observing traffic destined for servers that are participating as relays. More specifically, it has been reported that the probability of an adversary with IX-level access serving ASes appearing on both sides of a circuit is much greater than previously estimated [57]. This is due to traffic between the user and the entry relay, and also between the exit relay and website, passing through several routers existing in multiple ASes and IXs.

Another previous study conducted a real-world analysis of distribution of IXes on Tor entry and exit relay network paths to estimate the probability of the same IX or AS appearing at both ends of a circuit [34]. This involved running *traceroute* analysis from Tor clients to entry nodes, as well as from remote destination websites back to Tor exit nodes. All router points on the discovered route were then associated with an AS and an IX using available IP to AS number reference systems and IX mapping data. It was reported that one specific AS had a 17% chance of routing traffic for both ends of an established circuit, and a specific IX had a 73% chance [34]. This would ultimately allow completely passive analysis and correlation of traffic for large portions of the Tor network, and allow association of website interaction by a Tor user. This class of attack has a high level of sophistication due to the level of access required, and also due to the amount of processing power necessary to analyse the traffic and correlate traffic patterns. Organizations that do have the required access could make use of this capability to perpetrate such attacks themselves or share it with other parties. Changes to the Tor node selection algorithm have been suggested to reduce the feasibility of this attack by taking into account specific ASes and IXes on the paths between relay nodes [34].

It has been suggested that little attention has been given to application-level attacks against low-latency anonymous networks such as Tor [58]. A potential attack was presented in which attackers compromised multiple routers in the Tor network with only one web request to de-anonymise a client. However, there are several other requirements and caveats. Firstly, the target must construct circuits using compromised relays. Secondly, separate relay commands must be set up for each request and it's not clear whether this would work with simultaneous requests such as those over HTTP/1.1. The attack is noted as becoming far less effective if the client is participating in concurrent browsing activity, as this will dilute or completely hide the signal created. Furthermore, certain active circuits can be multiplexed over a single

---

[1] Individual and equal-sized 512 byte unit of Tor traffic [15]

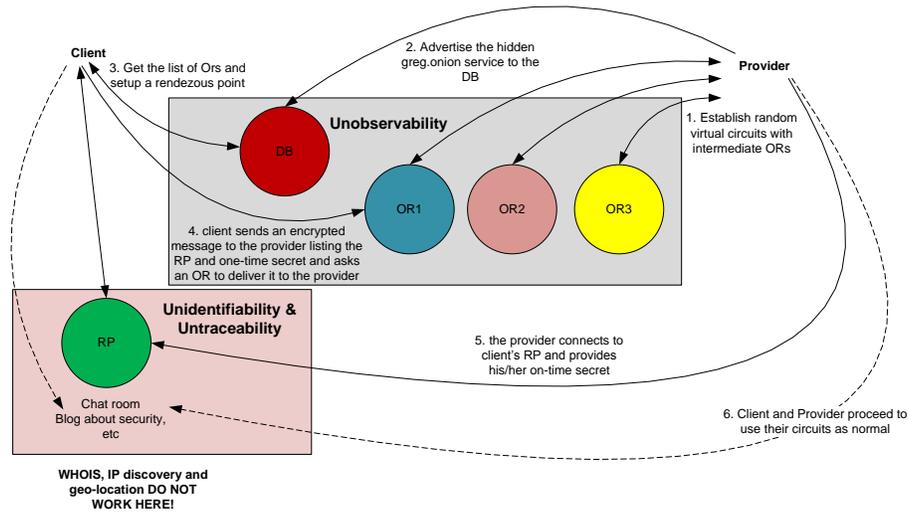

**Fig. 1.1** Hidden Services Protocol [15]

TCP stream with variations of the traffic pattern. Further developments by [20] identified that HTTP GET requests are typically encapsulated within a single Tor cell. The authors discussed how previous work on fingerprinting in single-hop encrypted connections such as SSH tunnels or openVPN was based on the fact that variations in web asset sizes are useful for allowing association of encrypted traffic to fingerprinted websites. However, since the introduction of HTTP/1.1 and the support for persistent connections handling concurrent requests, such attacks became largely ineffective due to overlapping web asset requests. As a result, attention quickly turned to packet sizes and timing. An important assumption in [20] is that HTTP GET requests exist in a single outbound cell from the client. The proposition is that by delaying these outbound packets by a small period between the client and the entry guard, overlaps between HTTP GET requests and server responses can be removed. This creates much cleaner traffic patterns for analysis and could overcome one of the possible limitations of [58], as already discussed.

### *1.2.2 Hidden Services*

The nature of Tor hidden services provides a number of challenges in collecting and analysing traffic, because by their very design the service host is intended to remain anonymous [15]. Many hidden services, or "deepnet" sites can be found using publicly available indexes. Such hidden services can then be fingerprinted for traffic analysis by a Tor client, but the wider challenge is in identifying and profiling sites that are not indexed in this way, or accessible by invitation only. There is research into attacks on hidden services in Tor [64, 28, 18], which, as stated, are as much of a problem for the provider of that service as the visiting user due to hidden service operators often wishing to remain anonymous. A brief overview of hidden service operation is provided in figure 1.1. For Tor networks in particular, running a hidden service comes with the greatest risk due to the ease with which an adversary can identify the service host [28]. Three separate attacks namely circuit classification, circuit fingerprinting, and website fingerprinting against Tor hidden services were described in [28] based on the assumption that it is possible to fingerprint the process of connecting to a hidden service. The ability to identify hidden services is significantly greater than association of normal user behaviour due largely to the much smaller number of hosts and typically more static content provided by hidden services compared to the open Internet [28]. Adversarial operators of hidden service directories may record the addresses of otherwise unadvertised hidden services to launch application level attacks [40].

There are many ways of covertly tracking a user's behaviour on the Internet such as cookies, server logs, and web beacons. Most PET technology will provide or support fit-for-purpose precautionary measures [44]. However, some attacks, especially against hidden services, may simply take advantage of the human factor. For example, configuration error is a large risk for inexperienced users of the i2p system [11] as well as Tor hidden service hosts. Hidden services can be erroneously hosted on a public-facing interface, or on a server that otherwise also hosts public information and gives away it's identity through private key fingerprints or other unique service information [11]. It is therefore recognised that PET systems are vulnerable to configuration errors.

### 1.2.3 Combining Methods

Little research attention has been drawn to the possibility that with sufficient resources, a powerful adversary could combine several documented attacks to further augment degradation of anonymity. Such combined methods are likely to have varying compatibility with each other but the potential end result would be some combination of an increased success rate, and a reduction in number of resources required as part of the attack itself. Considering the range of known attack methods against PET introduced above, we will now present a novel algorithm. Our algorithm combines a number of these previous attacks with a level of automatic adaptation to assist with reduction in un-traceability of Tor users. This can provide a fundamental pillar of a threat intelligence platform capable of identifying threats making use of PET.

## 1.3 Adaptive Traffic Association and BGP Interception Algorithm (ATABI)

Normally, the Tor client chooses three relay nodes to route encrypted traffic. Each node in turn then removes a layer of encryption, before the unencrypted traffic leaves the network towards its destination. Therefore, and as depicted in figure 1.2 step 1, server responses destined for the client are unencrypted until they re-enter the tor network, unless the website implements its own TLS encryption. Once the responses reach the exit relay (OR3) they are encrypted within the Tor network (step 2). Responses travelling between the entry relay (OR1) and the client are also encrypted as part of the established Tor circuit (step 3). For simplification, the following description assumes that an entity is interested in generating threat intelligence relating to undesirable use of PET on their infrastructure without being concerned about the potential repercussions of carrying out the methods described. A more pragmatic approach might be to collaborate with other service providers to perform wider passive analysis and reconnaissance, and agreeing about circumstances that collaborating entities are permitted to make BGP announcements for the purposes of association or investigation. Obviously operators could also share the necessary information with their collaborators without the need for BGP updates to gain additional network visibility.

As previously mentioned, the published research does not contain many detailed examples of combining known attacks against Tor, especially given the highly advantageous Man in the Middle (MITM) position provided to an AS- or IX-level entity. For instance, HTTP response traffic can be modified while in-transit with sophisticated regular expression (regex) search and replace methods [29, 36]. We propose that these or similar techniques can be used by an AS- or IX-level entity to manipulate server response traffic sent between websites and Tor exit nodes destined for the user. This will allow modification of HTTP responses sent back to the browser in order to directly affect or control client behaviour, while removing some of the obstacles to previous attacks such as having to compromise the website itself or for users to have JavaScript enabled in their browser. Obviously this kind of attack is considerably easier on clear text websites as opposed to HTTPS sites employing Secure Socket Layer (SSL) or TLS protocols, as in these cases the adversary would also have to gain visibility of server-side encrypted traffic. There are several known attacks against HTTPS that adversaries may be able to leverage to facilitate this, or they may simply have gained access to the private keys of a root or intermediate certificate authority [16, 37]. The AS or IX operator may instead wish to investigate watermarking techniques to avoid compromising HTTPS. The scope of this study is limited to testing HTTP traffic only assuming that HTTPS is either not enforced, or has been compromised.

Let us assume that an AS or IX operator has identified a website for which they wish to identify the users. If this entity has the capability of instigating a BGP interception attack, then they might begin by enumerating the list of all prefixes hosting Tor relays *that they do not already have visibility of*, or perhaps for locations that they believe to be likely sources of the traffic. With this information and the known prefix of the destination website, a BGP interception attack could be launched against all of these prefixes, thus getting complete visibility of the entry and exit traffic of interest. The challenge is then to associate the website traffic with the encrypted Tor entry traffic.

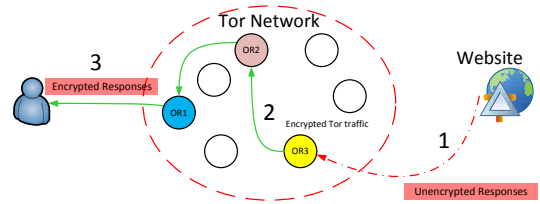

**Fig. 1.2** Server Responses in Tor

We present an algorithm 1 that considers whether or not an adversary already has visibility of their intended targets and performs BGP interception if necessary. The algorithm waits for a trigger condition such as a security alert, indicator of compromise, or manual initiation before saving all required network traffic to disk for analysis. Our algorithm consists of three components: BGP interception, MITM server response manipulation (using MITMProxy [36]), and a detection scheme. The full algorithm can be described as follows (see figure 1.3):

1. The IX-level adversary initiates a BGP interception against the subjects of interest, for instance a destination website. BGP interception can also be performed to gain increased visibility of client to entry-side traffic.
2. The adversary routes unencrypted web server response traffic through a manipulator in order to change the HTTP responses as required, for instance by inserting assets.
3. The traffic then enters the Tor network via the exit relay. From this point back to the client, responses remain encrypted as per the Tor implementation.
4. The adversary performs traffic analysis with the detection scheme on traffic destined for the client at point 2 (unencrypted) and point 4 (encrypted).

The expected outcome is that as routes start to converge, the adversary gains visibility of all traffic destined for that website as well as the proportion of Tor entry node traffic of interest. The MITMproxy configuration forces page responses to include large assets that increase traffic for users that the adversary desires. As mentioned, this attack has a high level of sophistication due to the type of access required by AS or IX Internet routing systems. However, for entities with that level of access, updating routing to flow through another system running some form of manipulation software would be straightforward. We believe that this is entirely achievable for a service provider or group of collaborating providers.

A number of assumptions based on previous research support the proposed algorithm. Firstly, we assume it is feasible to associate traffic behaviour to individuals if both sides of the connection can be observed [31, 22, 23]. Secondly, BGP routing attacks by an AS or IX operator can allow observation of large or specific proportions of Internet traffic [51, 57]. Finally, HTTPS is not always implemented, or may be vulnerable to a range of attacks, and can be vulnerable to watermarking techniques [16, 37].

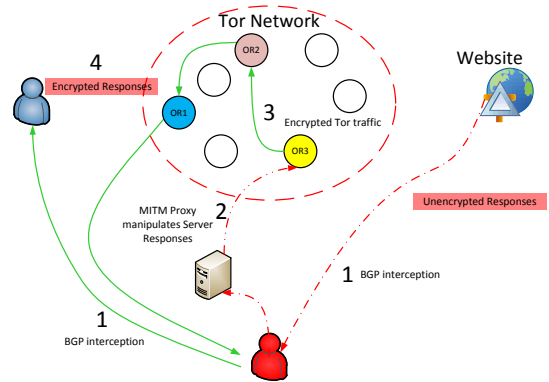

**Fig. 1.3** Adaptive Traffic Association and BGP Interception

### 1.3.1 BGP Interception Component

We consider the recently disclosed attack known as RAPTOR (Routing Attacks on Privacy in Tor) [51], which is a further development on previous works such as [57]. Such routing attacks are one example that could place a powerful adversary at great advantage, if they have access to core Internet routing infrastructure. These works describe three main assumptions that can be approached either individually or in combination to increase the exploitability of the system.

First, Internet routing is asymmetric in its nature. In other words, the path that an outgoing packet takes is not necessary the same as that of its reply, and visibility of only one direction of this traffic flow is required to analyse traffic. Therefore the attack surface and likelihood of exposure to an adversary performing a passive traffic association attack are greatly increased.

Second, BGP "churn" over time due to network changes or transient issues such as router failures increases the chances that a regular user of Tor will cross paths with a particular AS or IX, facilitating passive traffic analysis.

Finally, it is possible for an AS operator to make false BGP announcements in order to have traffic intended for another AS route through their own routers in an active attack, which positions themselves on a target circuit.

The active attack comes in two versions, hijack and intercept. The drawback of the hijack is that the traffic never reaches its intended destination and so the client experience will be interrupted, potentially raising the alarm, or causing a new circuit to be built. An interception on the other hand allows the traffic to continue to its destination and the client session remains active. There are three possible scenarios for the interception attack namely: 1) the adversary already monitors a website or exit node and targets a set of known Tor entry relays; 2) the adversary already monitors the client to entry relay side and targets known Tor exit relays (or a destination website); 3) the adversary targets both entry and exit traffic in order to perform general monitoring, data gathering and association. It should be noted that while the impact of a hijack to a user is obvious, there is no mention of the impact on user experience in terms of latency, or packet loss while the new routes propagate and converge during an interception attack.

Once this Man in the Middle position is achieved, the previous authors discussed the strong feasibility of traffic timing attacks, especially given the benefits of asymmetric routing. We propose that this MITM position can be put to further use to provide greater advantage by perpetrating further attacks in combination.

### *1.3.2 MITM Component*

Manipulating traffic is useful for affecting client behaviour. In a large set of similar candidate connections, the ability to affect those that the entity is interested in and separate them from the rest is obviously a valuable capability. Below, we demonstrate in a simple test that manipulation of server responses prior to entering the Tor network result in the expected client behaviour. We used the MITMProxy application to act as a reverse proxy for Apache HTTPd in order to replace:

```
</body>
```

With:

```
</body>
```

We suggest that an example of a real use case may instead need to simply forward traffic for all non-target destinations as required while pattern matching to manipulate only responses from the target website, and this is possible with MITM-Proxy filter options. We tested the "--anticache" option and found this to be effective against a caching server (varnish) in front of our test web server.

The output as observed in TORBrowser without MITMProxy was as follows:

```
curl —socks4a 127.0.0.1:9050 http://52.48.17.126/
<html>
<head>
</head>
<body>
<div align="center">
<p>Welcome home!!!</p>
</div>
</body>
</html>
```

And with MITMProxy running with search and replace on the body tag resulted to:

```
curl —socks4a 127.0.0.1:9050 http://52.48.17.126/
</html>
<head>
</head>
<body>
<div align="center">
<p>Welcome home!!!</p>
</div>
</body>
</html>
```

Note the additional image being included prior to the closing body tag, which is a 3.5MB JPEG but resized to one by one pixel so as not to be obvious. The image in this test also resided on the web server, but could just as easily be hosted elsewhere. However, hosting assets with another site may require alterations to the detection scheme.

### 1.3.3 Detection Scheme

Algorithm 1 includes the detection scheme, which is useful for fingerprinting connections and consists of a reduction function based on a set of filters applied in sequence. The goal is to eliminate client connection streams that are unlikely to be responsible for the observed server-side traffic, based on their meta-data. We propose that this may be useful in assigning a risk score to any client connections that remain after applying the detection scheme. Such scoring may be useful for later interpretation or alerting by Bayesian analysis or machine learning systems. Specifically, server-side sessions are fingerprinted by recording specific metrics including the start and end time of the connection ($SCt$ in Algorithm 1), total number of packets sent to the client ($tp$), average time distance between successive packets ($at$), total data sent ($td$), and total transmission time ($tt$). Similar metrics have been used in previous studies [22, 23, 31], however we believe our specific generation and treatment of these metrics is novel in its approach.

The same metrics are generated for all other observed connections. Once this fingerprinting is complete, reduction begins by first filtering for only connections that fit within the same time period as the server-side connection ($CCt$). Thereafter, remaining connections are filtered for similarity using the same metrics already discussed in turn, starting by leaving only those with similar total number of packets ($CCtp$), then similar average time between packets ($CCat$), similar total data received ($CCtd$), and finally, similar total transmission time ($CCtt$). Any remaining client connections are returned as possible candidates for the observed server-side connection in question. Moreover, our proposal is that average time distance between individual packets can act as the filter, followed by similar total amount of data sent, and then finally similar total transmission time.

The detection scheme is adaptive in the sense that an initial tolerance is set and if no candidates are found, the filter is performed again after automatically increasing the tolerance. In our initial tests, we selected a base interval of $\pm 1\%$ for increasing tolerances. This proved effective in our tests, but is adaptable and may improved in further work. This occurs for each of the discussed metrics in turn. For example, the

---

**Algorithm 1** Traffic Association

---

1: **function** BGPINTERCEPT(*target*)
2:     Initiate BGP Interception against *target*
3:
4: **function** COMPUTESTATS(*connection*)
5:     $tp$ = Total number of packets in *connection*
6:     $at$ = Average time between packets in *connection*
7:     $td$ = Total data sent in *connection*
8:     $tt$ = Total duration of *connection*
9:     **return** $tp, at, td, tt$
10:
11: **if** Target website traffic not visible **then** BGPIntercept *website*
12:
13: **if** Suspect clients traffic not visible **then** BGPIntercept *clients*
14:
15: **while** $trigger \neq 1$ **do**
16:     Check for trigger condition
17:
18: **while** $ServerConnection = active$ **do**
19:     Initiate MITM manipulation of server responses
20:     Save all network traffic to disk
21:
22: **for all** ServerConnections **do**
23:     $SCt$ = Timeframe of *ServerConnection*
24:     $SC$ = ComputeStats *serverconnection*
25:     **return** $SC$
26:
27: **for all** ClientConnections **do**
28:     $CC$ = ComputeStats *clientconnection*
29:     **return** $CC$
30: **for all** $SC$ **do**
31:     Set initial tolerances
32:     $CCt$ = list of $CC$ where $CC$ packets are in the same time frame as $SCt$
33:
34:     $CCtp = CCt$ where $CCt$ has similar total number of packets as $SC$
35:     **if** $CCtp = 0$ **then** Increase tolerance by $\pm1\%$ and repeat
36:     **if** tolerance $\geq$ max tolerance **then** Stop
37:
38:     $CCat = CCtp$ where $CCtp$ has similar average time between packets as $SC$
39:     **if** $CCat = 0$ **then** Increase tolerance by $\pm1\%$ and repeat
40:     **if** tolerance $\geq$ max tolerance **then** Stop
41:
42:     $CCtd = CCat$ where $CCat$ has similar total data sent as $SC$
43:     **if** $CCtd = 0$ **then** Increase tolerance by $\pm1\%$ and repeat
44:     **if** tolerance $\geq$ max tolerance **then** Stop
45:
46:     $CCtt = CCtd$ where $CCtd$ has similar total transmission time as $SC$
47:     **if** $CCtt = 0$ **then** Increase tolerance by $\pm1\%$ and repeat
48:     **if** tolerance $\geq$ max tolerance **then** Stop
49:
50:     **return** $SC, CCtt$

---

check for total packets sent will increase from $\pm 5\%$ by $\pm 1\%$ until at least one candidate is returned, up until a maximum tolerance level. Maximum tolerances were chosen simply to avoid the algorithm continuing indefinitely until candidates are returned that are highly unlikely to be those responsible for the server-side connection of interest. Any candidates found are passed to the next step in the detection scheme. The starting tolerance, incremental amount, and maximum tolerance are all configurable and were chosen by the present authors during testing as they proved to be effective with our data. We anticipate that future research will allow tolerance values to be set automatically by machine learning systems based on measurable factors including network conditions such as latency, packet loss, and jitter.

We will discuss experimental results following initial testing of the detection component of the algorithm, and offer a discussion of current design and performance. We use several key performance indicators (KPI's) used in typical binary classification systems to evaluate the effectiveness of the detection scheme. Our rationale for choosing these KPI's was due to the similarity to clinical studies in which a test is evaluated based on its ability to correctly diagnose members of a population with a condition, while keeping false positives to a minimum. KPI's include sensitivity (*se*, percentage of target clients correctly identified as candidates for a server-side connection), specificity (*sp*, percentage of non-target clients who are correctly excluded from the list of candidates), false positive rate (*fpr*, percentage of non-target clients who are incorrectly included in list of candidates), and false negative rate (*fnr*, percentage of victim clients who are incorrectly excluded from the list of candidates). For completeness, we also report the positive predictive value (*ppv*, percentage of included candidates who were truly responsible), and negative predictive value (*npv*, percentage of excluded clients who were truly not responsible).

## 1.4 Experimentation and Results

### 1.4.1 Experiment Setup

We used shadow [26] to create simulated networks and generate PCAP data for analysis. Shadow was used due to being readily available, easily accessible, multi-threading capability, simulation of transient network issues, its use of real Tor code, and the fact that it is still actively maintained [49]. Shadow and shadow-plugin-tor were built from the project source code available on github. The tgen plugin was used for the purposes of generating traffic on the network. This plugin makes use of graphml XML files that control the behaviour of a particular client such as which servers to download from, how large a download should be, how many downloads to complete, and how long to pause between each download. The latest stable release of Tor code (0.2.7.6 - 2015-12-10, at the time of experimentation) was used with shadow-plugin-tor.

In sizing our simulation network, we referred to [19] in which two separate experiments were conducted as follows; First, to test the feasibility of an attack, the authors implemented a small scale experiment consisting of 20 relays, 190 web clients and 10 bulk clients. Secondly, A larger network was also constructed by the authors based on the work of [27] consisting of 400 relays, 3000 clients, and 400 servers. Our approach was similar, implementing small networks for speed of simulations in order to debug any problems, and to develop and test the detection scheme. We then created a larger network resulting in more time-consuming simulation andgreater data generation for our testing.

AWS EC2 instances were used for running our simulations. With our time and budgetary constraints in mind, an m4.2xlarge AWS EC2 instance of 8 cores and 32GB memory was used to run simulations on a network containing 4 authority servers, 400 relays, 1000 clients, and 400 web servers. Generation of the network topology was based on tor metrics and server descriptors for the month of April 2016 (2016-04-30).

The 1000 clients included 989 similar web clients all set to download a 350KiB file every 60 seconds, and 10 bulk clients set to download a 5MiB file repeatedly without pausing. There were also two 50KiB, 1MiB, and 5MiB clients downloading every 60 seconds. The actual downloaded data is arbitrary and randomly generated by shadow in order to provide simulation data. To simulate increased traffic for a victim (e.g. following a MITM attack) a single default web client as provided by shadow was set to download a 2MiB file.

Due to the fact that shadow records session information for all clients, servers, and relays, it is straightforward to confirm from simulation output which client was actually responsible for a particular server connection. The process for running a particular test of the detection scheme was automated and involved the following steps:

- Inspect the web client log file for a successful download and take note of the time of the download and name of the destination web server.
- Inspect the web server log file for the correct download (which shadow helpfully confirms in the logs) and take note of the connecting exit relay session tuple (IP and port number).
- Run the algorithm against the given web server PCAP file and record connection tuple information of all clients' PCAP data in order to detect candidates for responsible clients.

The detection scheme is not currently multi-threaded but result generation for detection performance was scripted and split between 16 cores on a c3.4xlarge EC2 instance with 30GB memory, and with 9000 Input Output Operations Per Second (IOPS).

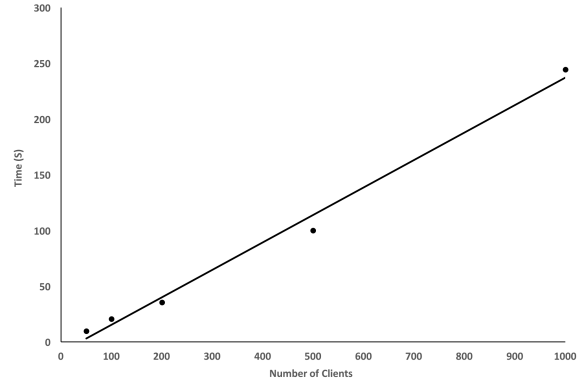

**Fig. 1.4** No. Clients vs Detection Run Time (s)

**Table 1.1** Single Detection Result Matrix

|  | Test Result | | |
|---|---|---|---|
| Client | Target | Non Target | |
| Target | $(tp = 1)$ | $(fn = 0)$ | $(t = 1)$ |
| Non Target | $(fp = 2)$ | $(tn = 197)$ | $(nt = 199)$ |
| Total | 3 | 197 | $(n = 200)$ |

### 1.4.2 Evaluation Criteria

Figure 1.4 illustrates a linear increase in average runtime of the detection scheme against a single session to identify a specific client amongst a group of 50, 100, 200, 500, and 1000 clients. This is indicative of the one to many relationship between a target connection on a server, and the variable number of potential candidates.

The detection scheme was tested by implementing in python against every client in turn within a particular set of simulation data and the detection performance KPI's were calculated. For a given detection attempt in a network of $n$ clients, the total number of targets $t$ will only ever be 1, with all other clients being non-targets ($nt = n - t$). This is because there is only ever one client that was truly responsible for a specific server-side connection. Using a diagnostic classification matrix, the outcome for a detection attempt contains values for true positives $tp$, being either 1 or 0; false negatives $fn$, (being the inverse of $tp$ given that there is only ever one true target); false positives where $0 \leq fp \leq (nt - 1)$; and true negatives $tn$ where $0 \leq tn \leq (n-1)$ and $tn = nt - fp$. A typical output from our testing is reflected in table 1.1.

Performance KPI's are all percentages in the range $0 \leq KPI \leq 1$. Sensitivity $se$ is calculated as $se = \frac{tp}{t}$ and will always be either 1 or 0 as there is only one target. Specificity $sp$ is calculated as $sp = \frac{tn}{nt}$. False positive rate $fpr$ is calculated as $fpr = \frac{fp}{nt}$ or $1 - sp$. False negative rate $fnr$ will always be 0 or 1 and is calculated as $fnr = \frac{fn}{t}$ or $1 - se$. Positive predictive value is $ppv = \frac{tp}{tp+fp}$ and Negative predictive value is $npv = \frac{tn}{fn+tn}$.

**Table 1.2** Detection Performance Comparison

| | Test | | | | | Test A Victim |
|---|---|---|---|---|---|---|
| | A | B | C | D | E | |
| *se* | 0.016±0.127 | 0.558±0.497 | 0.485±0.500 | 0.279±0.449 | 0.008±0.090 | 1.000 |
| *sp* | 1.000±0.001 | 0.981±0.014 | 0.986±0.010 | 0.993±0.006 | 1.000±0.000 | 0.999 |
| *fpr* | 0.000±0.001 | 0.019±0.014 | 0.014±0.010 | 0.007±0.006 | 0.000±0.000 | 0.001 |
| *fnr* | 0.984±0.127 | 0.442±0.497 | 0.515±0.500 | 0.721±0.449 | 0.992±0.090 | 0.000 |
| *ppv* | 0.012±0.100 | 0.073±0.186 | 0.075±0.192 | 0.065±0.181 | 0.008±0.090 | 0.500 |
| *npv* | 0.999±0.000 | 1.000±0.001 | 0.999±0.001 | 0.999±0.000 | 0.999±0.000 | 1.000 |
| Victim Detected: | True | True | False | False | False | True |

### 1.4.3 Results

We ran the detection scheme against all clients five times for the largest simulation data, varying the percentages of tolerance in order to identify whether there was an optimised tolerance for the different metrics. Tolerances for each metric began at ±5% in every test and as per the algorithm, increased by ±1% until candidates are returned or until the maximum tolerance is reached. These results are presented in table 1.2. The tolerance values tested were (number of packets, average packet time, total data, total time): 52,32,2,1 (A); 100,50,25,5 (B); 50,50,25,5 (C); 50,50,10,5 (D); 25,25,5,1 (E)

The results table also includes a column showing detection performance KPI's for the MITM victim during Test (A). The first test (A) used tolerance values tailored to achieve the best detection of the correct MITM victim target. Subsequent tests were performed to quantify detection KPI performance in general, and also tested against the MITM victim. Test (B) allowed a maximum tolerance of ±100% for similar number of packets and resulted in larger numbers of candidates to be included for classification according to the other metrics. Test results show how the performance KPI's are affected by changing the maximum tolerance values of the algorithm. Test (A) clearly performs poorly overall, however is highly successful in terms of true positive and false positive rates. This might be expected as test (A) was designed for detection of the MITM victim.

With these maximum tolerance values set, the correct target was identified only 16 times out of 984 total attempts, while detection performance of the MITM victim client with these tolerance values was very high. Detection of the MITM victim was achieved with a sensitivity of 1 (average for all other detections against non-victims = 0.016±0.127), specificity of 0.999 (average = 1±0.001) and a false positive rate of 0.001 (average = 0±0.001).

Test (B) correctly identified the MITM victim, and also correctly identified other targets with a sensitivity of 0.558 ±0.497 and false positive rate of 0.019±0.014.

Detection performance whilst searching for the victim with these tolerances had a sensitivity of 1 and $fpr$ of 0.003, which indicates an improvement over detection of non-victims.

Tests (C,D,E) all failed to identify the MITM victim and achieved decreasing detection performance KPI's as per table 1.2.

## 1.5 Discussion

Our tests indicate a poor detection performance for any client while using maximum tolerances for total number of packets, average inter-packet time, and total data sent, and lower maximum tolerance for total connection time (Test (B). While the MITM victim was detected with these tolerances, three false positives were also identified as candidates. With fine tuned tolerances for the MITM victim in Test (A), detection performance for other targets is poor, while the victim detection was successful with just 1 false positive. This suggests that getting the tolerances right is critical to success. Furthermore, with further development and testing, tolerances could be automatically defined based on external factors such as network conditions or factors that the adversary can control, such as the size or number of inserted assets. It may also be worth investigating the splitting of individual connections into sections that can be individually fingerprinted with the same metrics. The relationship between sections when comparing client to server-side traffic may yield improved detection rates thanks to temporal variations in network conditions. We note that the runtime performance of the detection scheme is not currently scalable, and that there are opportunities for further efficiencies to be added. Machine learning in particular is of interest as a solution for eliminating candidates when performing the traffic association element.

A general strength of this algorithm is that it does not require Tor nodes to be injected or compromised. The algorithm is simplified and only requires average and total statistics to be calculated for some individual flow meta data for comparison. By using average data for connections rather than considering all inter-packet timing data for comparison, minor transient network issues such as short-term jitter and packet loss are expected to have a reduced impact on detection, however false positives may be more of a problem when larger data sets are available. Shadow allows for random and transient variables such as bandwidth, computing power, packet loss, and jitter to be simulated in order to mimic real network connections and with these default measures in place, our initial results are positive.

The application layer component of this algorithm relies on visibility of unencrypted traffic between the exit node and the destination, in contrast to watermarking techniques, which operate purely by embedding a signal into the packet timing. Previous attacks such as SWIRL are also blind, meaning that as long as the watermarker and detector both know the watermarking secret, no other data needs to be synchronised between them. Our attack requires that the metadata descriptors of the

entry- and exit-side connection be shared with whichever system is performing the algorithm, at less than 150bytes per connection.

The attack also assumes that the client is only participating in activity on one destination website. Because the onion proxy multiplexes all outgoing connections through a single connection to an entry guard, if several client activities are concurrent then the algorithm will not currently deal with this. However, if we assume that in most cases users of Tor will be browsing one page at any one time, the time slicing of the client side multiplexed connection would in that case only include the relevant data pertaining to the observed exit-side traffic.

We currently make no attempt to remove Tor control cells from the client side connection data. The tolerance level in the algorithm helps to account for this, but a future improvement to the algorithm might wish to take this into account as discussed in [20]. Total data received by the client in a particular window could simply be indicative of client bandwidth limitations. Therefore, for a given time window of the exit-side traffic, there is a risk that other Tor users with similar bandwidth restrictions could appear in the set of candidates, even if they are accessing completely different sites. However this still allows for correlation with the exit-side traffic thanks to the inter-packet timing and total data sent for individual connections. We also note that compared to watermarking techniques, the performance impact of embedding large assets would perhaps be more noticeable in text only applications, image-sparse forums, or bulletin board systems.

### 1.5.1 Use Cases

It has previously been presented that a small number of very large ASes are naturally in a position to see at least one end of a circuit due purely to their size [51]. Ten ASes were shown to have visibility of at least 50% of all Tor circuits, with some providers seeing over 90%. This would place these providers in a prime position to generate threat intelligence data points based on traffic analysis, or to initiate BGP updates to hijack or intercept traffic during investigations. The initial threat intelligence gathering could lead to triggering a BGP routing update under certain conditions, such as a security alert. It is worth noting that there are numerous entities and organisations that monitor BGP activity on the internet and report suspected interception incidents [4]. As such, anomalies are generally reported in forums and news sites fairly quickly when they do occur. However, this does not imply that organisations will not carry out such methods anyway, only that they are more likely to be detected, reported and discussed.

We imagine two preliminary use cases for our algorithm during investigations. Firstly, an adversary may have little or no idea of the location of the sources of traffic destined for a particular website. The objective here would be a wide BGP interception attack against all traffic that the adversary does not already have visibility of. This would be a large-scale attack and require significant resource for injecting the HTML into responses and processing the traffic to identify the sources responsible. An alternative to reduce the overhead would be to engage in a number of smaller attacks iterating through multiple source locations. In this case the BGP interception attacks would be targeted at the smaller IP ranges in sequence together with the destination if required. As discussed, another alternative would be if AS or IX operators participate in threat information exchange, or agree terms under which such BGP routing updates are acceptable.

Secondly, the adversary may have a good idea of where the source of traffic is coming from, for instance in a criminal investigation where a suspect is believed to reside in a particular area. In this case, the adversary may be fortunate enough to already have visibility of the required IP ranges thanks to asymmetric routing but if not, then would only have to perpetrate a much smaller BGP interception attack based on the suspect's location.

While our presented algorithm applies to Tor clients accessing public Internet sites, similar techniques could be applied to Tor hidden services. For instance, if an operator was to first identify a remote hidden service, then they could repeatedly make custom requests to it during a widespread BGP interception attack (or working with collaborators) while running the detection scheme. If detection of the hidden service were to prove successful, then BGP interception and detection could be used to identify clients of the hidden service, or the adversary could target the hosting system directly in other ways, leading to further attacks.

### 1.5.2 Proposed Defences

To mitigate against the HTML injection component, Tor nodes could consider disallowing port 80/HTTP in their exit policy. This may be strongly advised in any case considering the typical Tor user's privacy concerns as well as the wider industry moving towards HTTPS everywhere, and HTTP Strict Transport Security (HSTS). If blocking port 80 is too problematic for user experience or website functionality, then the only other option is for website operators to correctly implement HTTPS and make use of HSTS. Users that care deeply about their privacy should insist on using HTTPS websites and avoid the use of HTTP sites for accessing or sharing sensitive information. This will still not help if the adversary has the ability to compromise HTTPS, but would make success for the adversary significantly more difficult, or force the use of other methods such as watermarking.

The countermeasures suggested by the authors of RAPTOR [51] would all still apply to the routing component of the current algorithm and therefore deserve a mention. Monitoring was proposed to raise awareness of the possibility of increased traffic visibility due to asymmetric routing, BGP churn, or routing attacks, and to notify clients when a potential degradation of anonymity is identified. BGP and traceroute monitoring were both suggested and tested with successful results reported.

It was also noted that traditional countermeasures that manipulate packet sizes or timing come with a significant latency impact and so would remove one of the main benefits of Tor. A number of alternative countermeasures were proposed, including incorporating traceroute and AS lookups as part of intelligent selection of relays to build circuits that avoid routes crossing the same AS multiple times. However, this may result in lower overall entropy in Tor circuit selection and greater probability of adversaries with large numbers of injected or compromised relays appearing on both ends of a circuit.

Another suggested mitigation was that Tor relays should advertise prefixes of /24 to reduce the capability of BGP hijack or interception attacks. Adversaries could still launch attacks by advertising the same prefix, however the impact of the attack would be more localized to the adversary and more widespread redirection of Internet traffic would not take place. To further mitigate this reduced capability, it was suggested that clients could favour guard relays that are closer in terms of number of AS hops, but that this could reveal probabilistic information about clients, and requires further investigation. The final suggested mitigation requires that the wider Internet moves towards secure inter-domain routing, however this has been slow to get off the ground.

## 1.6 Conclusion and Future Work

This work has presented a novel algorithm (ATABI) against the Tor network based on the combination of previously reported HTML injection techniques with BGP routing attacks and a detection scheme. This algorithm could form a fundamental pillar of a PET-capable Cyber Threat Intelligence Management platform. A simplified version of the MITM component was perpetrated against a basic web page to insert a large hidden asset at the bottom of the HTML. This change in the returned HTML was observed in the response received by the Tor client. The suggested detection algorithm yielded positive results in initial tests on data generated by the shadow simulator for Tor. Detection performance of an MITM victim with large tolerances was good with a $fpr$ of 0.003, while average general detection performance of all clients without MITM traffic manipulation was poor with sensitivity of $0.558\pm0.497$ and $fpr$ of $0.019\pm0.014$. By tailoring the tolerances for the MITM victim and running against all clients, general performance was greatly reduced with average sensitivity of $0.016\pm0.127$ and average $fnr$ of $0.984\pm0.127$, while sensitivity when searching for the victim was 1 with a $fpr$ of 0.001.

In future research we intend to evaluate whether temporal patterns manifested throughout the duration of a session can assist detection by splitting sessions and fingerprinting each section. We hope to provide more practical examples of routing attacks facilitating a MITM position and consider other MITM technologies or development of a dedicated lightweight application. With further efficiencies such as multi-threading for an individual attack, performance could be increased to operate with larger data sets. Increased performance may also be achieved by using GPU, or distributed computing. Combining the IX-level advantage with other previous concepts such as watermarking schemes, sending custom requests to hidden services, or delaying GET requests but at the exit side rather than the client side are also of interest. We also plan on performing further simulations and experiments to produce more data in order to further optimise the tolerance levels for each connection metric, and provide more data for potential testing and implementation of machine learning. Given the large number of CTI ontologies available, we would see benefit in developing our algorithm to be platform agnostic. A worthwhile project would be the creation of a PET-capable Cyber Threat Intelligence Management Platform that can interoperate with a wide range of CTI ontologies, with a view to driving standardisation, and with our algorithm as a fundamental pillar.

We believe that this variety of opportunities to perpetrate previously known attacks against PET more effectively in combination with AS- or IX-level access is likely to gain more research attention in the future. This likelihood is augmented especially considering the challenge of attribution in CTI for darknet-based threat activity.